\documentclass[12pt,oneside]{article}

\usepackage{amsfonts,amsmath,amssymb,epsfig}
\begin{document}
\title{\vspace{-40pt} What if Quantum Gravity is \\ ``just'' Quantum Information Theory?}
\author{Aron C. Wall}

\maketitle

\begin{abstract}
I suggest the possibility that holographic quantum gravity is, in some sense, equivalent to quantum information theory.  Some radical implications would follow.  First, the theory of quantum gravity should have no adjustable coupling constants, similar to string theory.  Thus, all complete bulk theories of quantum gravity are dual to each other.  By setting up an appropriately entangled state, it should be possible to find wormholes connecting any two quantum gravity theories (e.g. string theory and loop quantum gravity).  Secondly, if we represent space at one time as a tensor network, then dynamics is automatically encoded via gauge-equivalent descriptions of the boundary state.  This would appear to imply, contrary to semiclassical expectations, that a closed universe should have only one state.
\end{abstract}

\vspace{50pt}

\begin{centering}
Prepared for publication in: \linebreak
\textit{Proceedings of the 28th Solvay Conference on Physics:\linebreak The Physics of Quantum Information}\linebreak ed.~David Gross, Alexander Sevrin, Peter Zoller  \linebreak World Scientific Publishing Co., Singapore, 2023.

\vspace{12pt}
Conference held in Brussels, May 19-21, 2022.

\vspace{12pt}
~~~~~~~~~~~~Additional references have been added to the arxiv version.
\end{centering}

\newpage
\large


I want to talk about an idea which has been discussed recently in the holography community, but whose radical implications have perhaps not been fully appreciated.  This idea might be wrong.  But I think it is important to take such ideas with maximal seriousness---not because they are necessarily right; but because that way you find out faster if they are wrong.  If we make too many excuses for our ideas, and say we are just studying ``toy models'', then we lose the ability to discover that our principles don't work.

The idea is that quantum gravity (QG) in some sense reduces to, or is morally equivalent to, quantum information (QI) theory.  I decided to talk about this because there are a lot of quantum information theorists here---and if I were working on quantum gravity without knowing it, I would want someone to tell me!

(Some researchers associated with developing this idea are L. Susskind \cite{Maldacena:2013xja,Susskind:2017ney}, B. Swingle \cite{Swingle:2009bg}, T. Takayanagi \cite{Nozaki:2012zj,Miyaji:2015yva}, and M. Van Raamsdonk \cite{VanRaamsdonk:2010pw}.  I'm only mentioning people who aren't here at this Solvay conference---if you are here, then you know who you are.)

To see how radical this ``QG = QI'' idea is, let's explore some of the implications:

1. The first thing I want to note is that there is essentially just \emph{one} quantum information theory (over the complex numbers $\mathbb C$).  Roughly speaking, it tells us about all possible ways that finite dimensional Hilbert spaces can be entangled with each other.

This suggests that there is also a unique theory of quantum gravity (at least, only one theory which is compatible with the holographic principle), with no adjustable free parameters (other than dynamically evolving fields).  That's a familar property from string theory, which \emph{also} has no adjustable coupling constants.  But it would also have to apply to any other approach to constructing holographic bulk theories.  Indeed, all such approaches would need to be dual to each other.

Let me state this in the most politically provocative way I can.  Consider ${\cal N} = 4$ Super Yang-Mills, a theory which is known to be dual to string theory in an Anti de Sitter background.  Suppose now that the loop quantum gravity (LQG) community gets their act together and figures out how to quantize a LQG model in AdS.  Then there are good reasons\footnote{See e.g. \cite{Marolf, Marolf:2008mg, Chowdhury:2021nxw}.} to think that this model is also dual to a CFT.  So now we have a stringy CFT and a LQG CFT.  Since these are quantum systems, nobody can stop us from considering states in which these two CFTs are entangled with each other.

\begin{figure}[ht]
\centering
\includegraphics[width=5.0in]{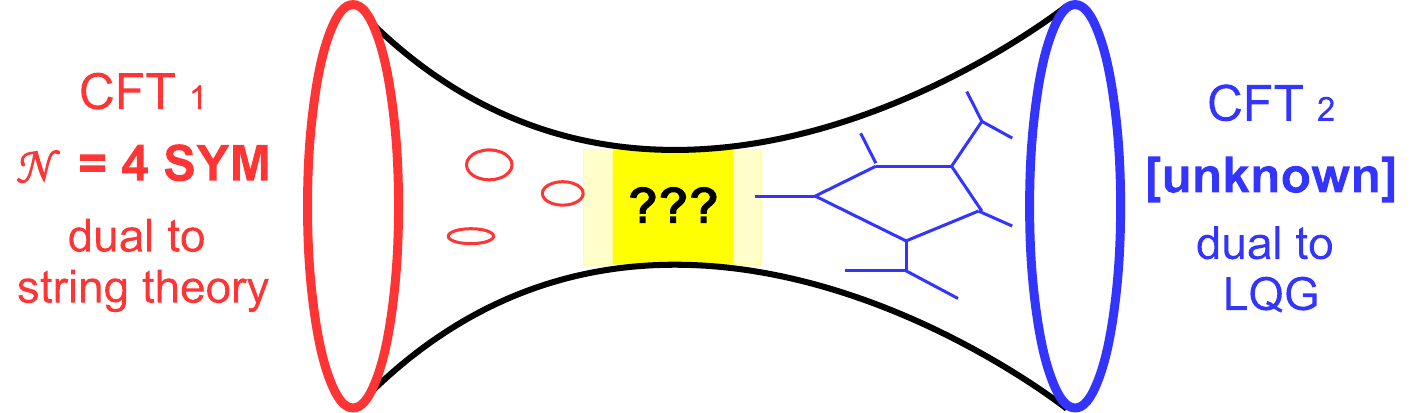}
\caption{Wormhole connecting string theory and LQG}\label{Wall1}
\end{figure}


Well, in the holography community, we believe that if two systems are sufficiently strongly entangled, this is equivalent to a wormhole (possibly a very complicated or long one) connecting the two sides (see Fig.~\ref{Wall1}).  So in the bulk, we would have a wormhole throat which somehow interpolates between string theory and LQG.  It seems like they would therefore be different dual descriptions of the same theory.

2. One way to try to make this idea a little more precise is using the concept of tensor networks.  As most people here know, a tensor network defines a state living on its boundary.  
\begin{figure}[ht]
\centering
\includegraphics[width=1.9in]{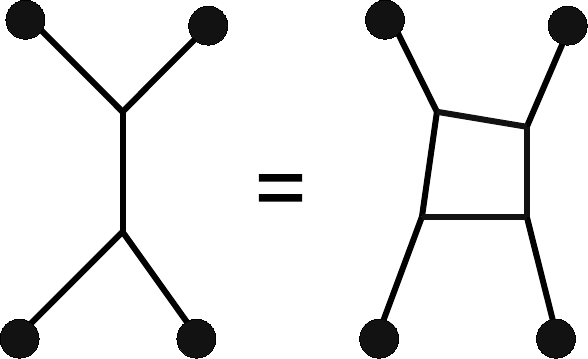}
\caption{Equivalence of Tensor Networks}\label{Wall2}
\end{figure}


Now I want to define two (quantum superpositions of) tensor networks as \emph{gauge-equivalent}, if they produce the same boundary state as each other.  For example, if you look at Fig.~\ref{Wall2}, there are 4 boundary sites, and the two tensor networks shown are gauge-equivalent if they are different ways of encoding the same entangled state, on the tensor product of these 4 boundary Hilbert spaces.

This is reminiscent of what happens in General Relativity (GR), where there is also quite a lot of gauge invariance.  Recall that in GR, local time evolution is also regarded as a gauge symmetry (since the time ``t'' is an arbitrary coordinate choice).  If QG = QI, then time evolution needs to somehow be a subset of this tensor network gauge invariance.  In other words, the ``future'' is just an alternative way of computing whatever boundary state the past computes.\footnote{For a continuum version of this same idea, for Cauchy slices of AdS/CFT, see \cite{Araujo-Regado:2022gvw}.}

Furthermore, on a black hole horizon you could consider the tensor network edges poking through the horizon.  Then we expect that the log of the dimensions of the tensor network edges should sum to the Bekenstein-Hawking entropy: $\text{Area}/4$ in Planck units.  (Although this is not in general going to be true away from the horizon.)

One particularly surprising consequence of the idea that QG = QI, is that it would imply that a spatially closed cosmology has only 1 allowed state.  (After all, a closed tensor network just evaluates a number, which lives in the trivial 1d Hilbert space.)  This differs from semiclassical expectations where one expects a great many possible states of a closed universe.  Whether this is a flaw in the idea, or whether holographic cosmology only makes sense if we live in a world with a spatial boundary, is worth pondering more deeply.

Let me end by rephrasing this idea in the form of a question:  Which cherished physical principles (unitarity, locality, etc.) are \emph{emergent} in the sense that they follow naturally from the quantum information paradigm, and which ones need to be \emph{postulated}?  If there are important physical restrictions on allowable spacetimes which go beyond the requirements of quantum information theory, then we had better identify what they are.  Thank you for your attention.

\normalsize

\subsubsection*{Acknowledgement} I am grateful to the Solvay Institutes for their hospitality.  The development of these ideas was supported by AFOSR grant FA9550-19-1-0260 “Tensor Networks and Holographic Spacetime”, and an Isaac Newton Trust Early Career grant.  I am also grateful for numerous conversations about emergent spacetime with Ted Jacobson, Don Marolf, Will Donnelly, Raphael Bousso, Juan Maldacena, Gon\c{c}alo Araujo-Regado, Rifath Khan, Ronak Soni, and Vasu Shyam.

\end{document}